\documentclass[conference]{IEEEtran}
\usepackage{color}
\definecolor{mygreen}{rgb}{0,0.5,0}

\usepackage{listings}

\usepackage{cite}

\ifCLASSINFOpdf
\usepackage[pdftex]{graphicx}
\else
\fi
\ifCLASSOPTIONcompsoc
 \usepackage[caption=false,font=normalsize,labelfont=sf,textfont=sf]{subfig}
\else
 \usepackage[caption=false,font=footnotesize]{subfig}
\fi
\usepackage[usenames,dvipsnames,svgnames,table]{xcolor} 
\usepackage{xspace} 
\usepackage{multirow}
\usepackage{booktabs}
\usepackage{framed}
\usepackage{etoolbox}

\newcommand{\TRSM}{{\sc trsm}\xspace}
\newcommand{\POTRF}{{\sc potrf}\xspace}
\newcommand{\SYRK}{{\sc syrk}\xspace}
\newcommand{\GEMM}{{\sc gemm}\xspace}
\newcommand{\IDLE}{{\sc idle}\xspace}

\newcommand{\JUNO}{{\sc Juno}\xspace}
\newcommand{\ODROID}{{\sc Odroid}\xspace}
\newcommand{\DAG}{\emph{DAG}\xspace}

\newcommand{\botlev}{{\sc Botlev}\xspace} 
\newcommand{\BOTLEV}{\botlev}
\newcommand{\CATS}{{\sc CATS}\xspace}
\newcommand{\CPATH}{{\sc CPATH}\xspace}
\newcommand{\HYBRID}{{\sc HYBRID}\xspace}

\newcommand{\OMPSS}{OmpSs\xspace}

\newcommand{\BIG}{{\sc big}\xspace}
\newcommand{\LITTLE}{{\sc LITTLE}\xspace}

\newcommand{\fg}[1]{\textcolor{OliveGreen}{#1}} 
\newcommand{\br}[1]{\textcolor{BrickRed}{#1}} 

\newcommand{\Pzero}{{\sc PBotlev}\xspace}

\newcommand{\Pfs}{{\sc FS}\xspace}
\newcommand{\Pts}{{\sc TS}\xspace}

\newcommand{\Pone}{{\sc FS1}\xspace}
\newcommand{\Ptwo}{{\sc FS2}\xspace}
\newcommand{\PtwoP}{{\sc FS2'}\xspace}
\newcommand{\Pthree}{{\sc FS3}\xspace}

\newcommand{\Pfour}{{\sc TS1}\xspace}
\newcommand{\Pfive}{{\sc TS2}\xspace}
\newcommand{\Psix}{{\sc TS3}\xspace}

\begin{document}
%
\title{Energy efficiency optimization of task-parallel codes on asymmetric architectures}

\author{\IEEEauthorblockN{Luis Costero, Francisco D. Igual, Katzalin Olcoz and Francisco Tirado}
\IEEEauthorblockA{Departamento de Arquitectura de Computadores y Autom\'atica\\
Universidad Complutense de Madrid\\
Email: \{lcostero, figual, katzalin, ptirado\}@ucm.es}
}


%


\maketitle

\begin{abstract}
We present a family of policies that, integrated within a runtime task scheduler (Nanox),
pursue the goal of improving the energy efficiency of task-parallel executions with no 
intervention from the programmer. The proposed policies tackle the problem by modifying 
the core operating frequency via DVFS mechanisms, or by enabling/disabling the mapping 
of tasks to specific cores at selected execution points, depending on the internal status of the scheduler. 
Experimental results on an asymmetric SoC (Exynos 5422) and for a specific operation (Cholesky factorization) 
reveal gains up to 29\% in terms of energy efficiency and considerable reductions
in average power.\\

\textit{Task parallelism; runtime task schedulers; asymmetric architectures; energy efficiency; DVFS}
\end{abstract}

%
\IEEEpeerreviewmaketitle

\section{Introduction}
\label{sec:intro}

Asymmetric Multiprocessors (AMPs) are a class of heterogeneous parallel architectures in which
cores that implement different microarchitectures share a common ISA (Instruction Set Architecture) and,
possibly, a subset of memory resources. 
Typically, the available architectural heterogeneity is exploited pursuing
energy efficiency and performance restrictions on heterogeneous software environments.
%
%
One of the most popular implementations of AMPs is the big.LITTLE architectural paradigm
present in many ARM SoCs (Systems-on-chip), that combines a number of high performance ARM Cortex-A57/A15
\BIG cores with a (possibly different) number of energy-efficient ARM Cortex-A53/A7 \LITTLE cores.
Leveraging low-power architectures to the HPC (High Performance Computing) arena is one of
the main trends in the road towards the Exaflop barrier. Among them, ARM Cortex-A processors, and more
specifically, asymmetric SoCs based on this microarchitectural family, are nowadays on the spotlight as
the most promising architectures to achieve such a goal. 

However, increasing the heterogeneity entails a non-negligible impact on the
programmability of such platforms. In the last decade, task-parallel programming
models have emerged as an interesting solution that combines a correct orchestration of parallel 
programs and a reduced impact on the complexity of the parallel versions of existing or new codes. 
These models aim at casting a complete computation in terms of discrete pieces of code ({\em tasks})
with data dependences among them with the aid of task annotations provided by the programmer, and
rely on a {\em runtime task scheduler} (or just {\em runtime} in the following) that orchestrates the correct 
ordering of tasks execution as dependences are satisfied at run time.

The extension of these programming models and associated {\em runtimes} to heterogeneous architectures,
managing data coherency and data transfers among isolated memory spaces has been implemented in
a number of software efforts, together with techniques that drive to performance gains
in multi-core, many-core, accelerator-based and distributed-memory architectures. 
The necessary efforts to adapt these programming models to AMPs is also a topic of interest of recent
works~\cite{botlev,cpath,ashes16,Chen2014,Torng16}, pursuing the goal of boosting performance by correctly mapping critical tasks to the 
most appropriate element of the asymmetric architecture. These works complement energy-efficiency studies specifically targeting 
asymmetric architectures~\cite{Donyanavard16,Pricopi2013}.
However, the impact and possibilities of {\em task schedulers} in terms of improving energy efficiency 
of task-parallel implementations has not been previously studied in such a level of detail.
As of today, similar efforts, together with their impact on performance and energy efficiency, have not been ported or adapted
to AMPs. 

In this paper, we propose an extension of Nanox, the runtime task scheduler underlying
the \OMPSS~\cite{ompss} programming model that pursues the goal of reducing energy consumption with minimal impact on performance
and programmability.
We introduce a set of policies that modify both
task scheduling algorithms and frequency of operation of modern AMPs via DVFS depending on the
internal status of the task scheduler,
and evaluate their impact on both performance and energy efficiency on a Cholesky factorization (a widely used routine in many problems that arise in science and engineering, and illustrative of others DLA (dense linear algebra) implementations with similar features)
and an implementation of the big.LITTLE architecture (a Samsung Exynos 5422 SoC).

The rest of the paper is structured as follows.
Section~\ref{sec:runtime} reviews the state-of-the-art in modern task-parallel programming
models and their adaptation to asymmetric architectures.
Section~\ref{sec:policies} describes a number of energy-aware policies and mechanisms
that pursue an improvement in performance and energy-efficiency of Nanox on AMPs.
Section~\ref{sec:results} reports the impact of the aforementioned policies
in terms of performance and energy efficiency on the Exynos SoC.
Section~\ref{sec:conclusions} closes the paper with general remarks and 
future work.

%


\section{Runtime-based parallel execution on asymmetric platforms}
\label{sec:runtime}

A number of task-based programming models have previously proved to be an
efficient solution towards the exploitation of parallelism on multi-core,
many-core and heterogeneous architectures. In general, these models provide
a mechanism to annotate sequential codes and to indicate potential points
of parallelism, that is exploited at runtime by a task scheduler that takes
care of data dependences among tasks and a proper task-to-processor
mapping, typically improved by heuristics.  Among others, following the
path pioneered by
Cilk~\cite{Blumofe:Cilk}, efforts like
StarPU~\cite{Augonnet:2011:SUP:1951453.1951454},
Superglue~\cite{tillenius:superglue},
QUARK~\cite{icl:609}, Kaapi~\cite{Gautier:2013:XRS:2510661.2511383}, and \OMPSS~\cite{ompss}
pursue a common goal: extracting and exploiting task parallelism on modern
parallel architectures with minimal intervention of the programmer.

\OMPSS is one of the most widely accepted programming models nowadays. At a glance,
this programming model is based on the inclusion of directives ({\tt pragma}s) similar
to those used in OpenMP, that annotate specific sections of codes as {\em tasks}, that is,
minimum scheduling units to the available execution resources or processors. These annotations
include information about operands and their directionality (input, output and input/output). At
runtime, this information is handled by a {\em task scheduler} (named Nanox) that maps each task to the most
appropriate computational resource available as the inferred data dependences are satisfied.

\subsection{A driving example: the Cholesky factorization}

In the following, we employ the Cholesky factorization of a dense matrix as an illustrative 
example of the necessary modifications required by \OMPSS to extract and exploit the available
task parallelism in a specific operation.
%
Given a symmetric positive definite matrix $A$ of dimension $n \times n$, the Cholesky 
factorization decomposes it into $A = U^TU$, where the Cholesky factor $U$ is an upper triangular matrix.
Listing~\ref{lst:chol} sketches a C implementation of a blocked Cholesky implementation for a blocked matrix
$A$ composed of $s \times s$ blocks of dimension (block size) $b \times b$ each. Note that the routine decomposes
the global operation into a collection of basic kernels or fundamental operations, namely: 
{\tt po\_cholesky} (Cholesky factorization of the diagonal block); 
{\tt tr\_solve} (solution of a triangular system);
{\tt ge\_multiply} (general matrix-matrix multiplication);
and 
{\tt sy\_update} (symmetric rank-$b$ update).

\begin{lstlisting}[float=th!,frame=lines,caption=C implementation of the blocked Cholesky factorization.,label=lst:chol]
void cholesky (double *A[s][s], int b, int s) {
   for (int k = 0; k < s; k++) {
      // Cholesky factorization
      // (diagonal block)
      po_cholesky (A[k][k], b, b);   
                                    
      for (int j = k + 1; j < s; j++)  
         // Triangular system solve.
         tr_solve (A[k][k], A[k][j], b, b); 

      for (int i = k + 1; i < s; i++) {
         for (int j = i + 1; j < s; j++)
            // Matrix-matrix multiplication.
            ge_multiply (A[k][i], A[k][j], 
                         A[i][j], b, b);     
         // Rank-b update.
         sy_update (A[k][i], A[i][i], b, b); 
      }
   }
}
\end{lstlisting}

\begin{figure}
\begin{center}
\includegraphics[scale=0.35]{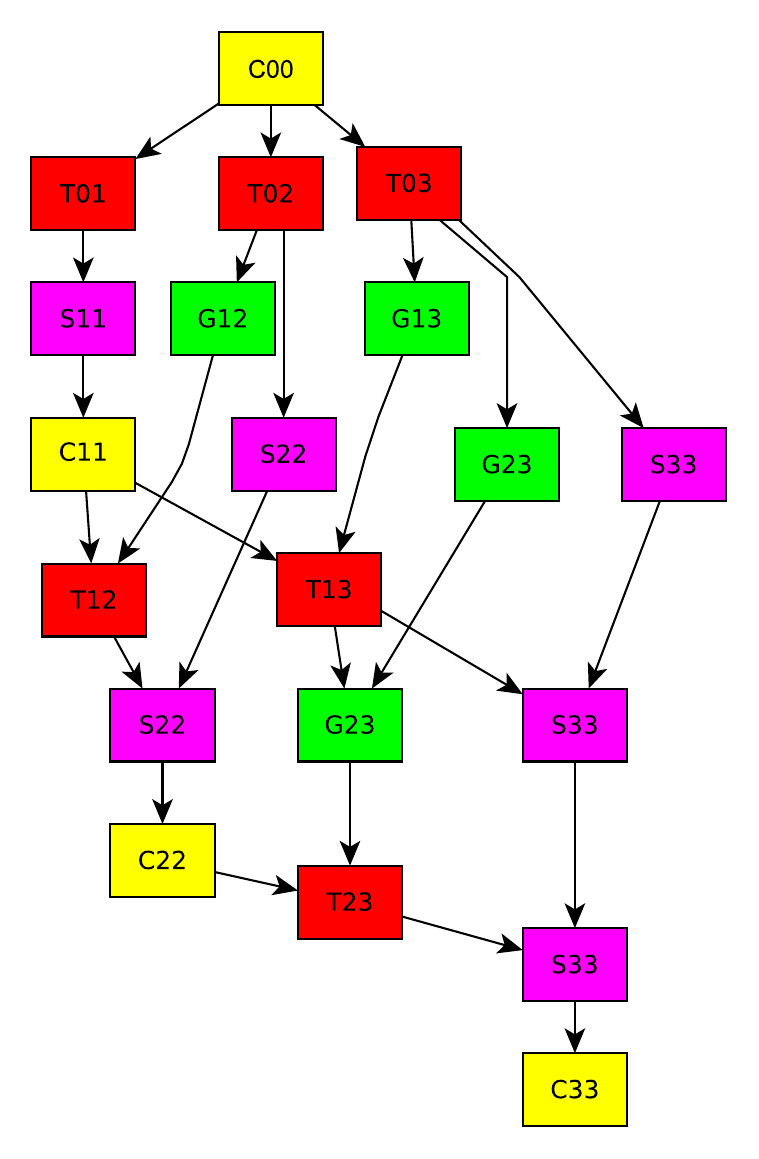}
\end{center}
	\caption{DAG with tasks and data dependences extracted from the application
of the code in Listing~\ref{lst:chol} on a matrix 
           with $4 \times 4$ blocks ({\tt s}=4). The labels in the nodes 
           specify the type of kernel/task as follows:
           ``{\sf C}'' for the Cholesky factorization;
           ``{\sf T}'' for the triangular system solve; ``{\sf G}'' for the matrix-matrix multiplication, 
and ``{\sf S}'' for the symmetric rank-{\tt b} update.  The subindices (starting at 0)
           specify the submatrix updated by the corresponding task.}
\label{fig:dag}
\end{figure}

These are the fundamental parts of the overall computation, or {\tt tasks}. Obviously, provided each
task is internally executed in a sequential fashion, the aforementioned code would not extract any further
level of parallelism. Listing~\ref{lst:chol_tasks} includes the necessary modifications in the definitions
of each task in order to exploit the \OMPSS programming model and, thus, to extract task parallelism in a transparent
manner. Note how each task is annotated with the corresponding {\tt pragma omp task} directive, including the
directionality of each operand involved in the computation. At runtime, the invocation of each task in Listing~\ref{lst:chol}
is intercepted by the runtime task scheduler (Nanox), that dynamically builds a DAG (Directed Acyclic Graph) as the one
shown in Figure~\ref{fig:dag}, including tasks (nodes) and data dependences among them (edges). Only when all data
dependences for a given task are satisfied, the runtime dispatches that task to an available processor, effectively
exploiting task parallelism.

\begin{lstlisting}[float=th!,frame=lines,caption=Annotated tasks for the blocked Cholesky factorization.,label=lst:chol_tasks]
#pragma omp task inout([b][b]A)
void po_cholesky (double *A, int b, int ld) {
 static int        INFO = 0;
 static const char UP   = 'U';

 // LAPACK Cholesky factorization
 dpotrf (&UP, &b, A, &ld, &INFO);  
}

#pragma omp task in([b][b]A) inout([b][b]B)
void tr_solve (double *A, double *B, int b, int ld) {
 static double     DONE = 1.0;
 static const char LE   = 'L', UP = 'U', 
                   TR = 'T',   NU = 'N';

 // BLAS-3 triangular solve
 dtrsm (&LE, &UP, &TR, &NU, &b, &b, 
        &DONE, A, &ld, B, &ld);    
}

#pragma omp task in([b][b]A,[b][b]B) inout([b][b]C)
void ge_multiply (double *A, double *B, 
                  double *C, int b, int ld) {
 static double     DONE = 1.0, DMONE = -1.0;
 static const char TR   = 'T', NT    = 'N';

 // BLAS-3 matrix multiplication
 dgemm (&TR, &NT, &b, &b, &b, 
        &DMONE, A, &ld, B, &ld, &DONE,  C, &ld);   
}

#pragma omp task in([b][b]A) inout([b][b]C)
void sy_update (double *A, double *C, int b, int ld) {
 static double     DONE = 1.0, DMONE = -1.0;
 static const char UP   = 'U', TR    = 'T';

 // BLAS-3 symmetric rank-b update
 dsyrk (&UP, &TR, &b, &b, 
        &DMONE, A, &ld, &DONE,  C, &ld);    
}

\end{lstlisting}

\subsection{Asymmetry-aware task schedulers}
\label{sec2:asymmetric-scheduler}

The design of efficient task scheduling algorithms on multi-core and
heterogeneous systems has been extensively studied in the past. 
Some of these works have been recently extended in order to accommodate
AMPs as the target platform. 
Examples of these efforts are \CATS~\cite{botlev} or \CPATH and
\HYBRID~\cite{cpath}, which aim at dynamically identifying which tasks belong
to the critical path of the DAG, assigning them to the fastest cores,
thus reducing the total execution time.

In \OMPSS the \CATS implementation is called \BOTLEV (\emph{Bottom level-aware
  scheduler}), and it has been used as a starting point for our work.
\BOTLEV dynamically detects the longest path of the DAG, assigning
those tasks that belong to it to the fast cores of the system. In order to
determine which tasks belong to the longest path, each task is internally annotated
with the longest distance between itself and a leaf task. Each time a new
task is inserted into the DAG, all of its predecessor nodes in the graph
are updated only if the longest path increases; proceeding this way, the 
longest distance between each task and a leaf node is always updated.

When a task becomes ready for execution, it is classified as critical 
or non-critical based on the annotated distance: if it belongs to the longest known path, it is
stored as a {\em critical} task. Ready critical and ready non-critical tasks are
stored in two different priority queues sorted by its annotated
distances. When a core becomes idle, it retrieves a ready task depending on
the kind of core: \BIG cores execute ready tasks stored in the critical
queue, and \LITTLE cores retrieve tasks from the non-critical
queue. \BOTLEV enables work stealing for \BIG cores by default, allowing
\BIG cores to execute non-critical tasks if the critical-queue is
empty. Optionally, work stealing can be activated in a bi-directional fashion.

\subsection{Target asymmetric architectures}

\begin{figure}
  \centering
  \includegraphics[width=0.6\columnwidth]{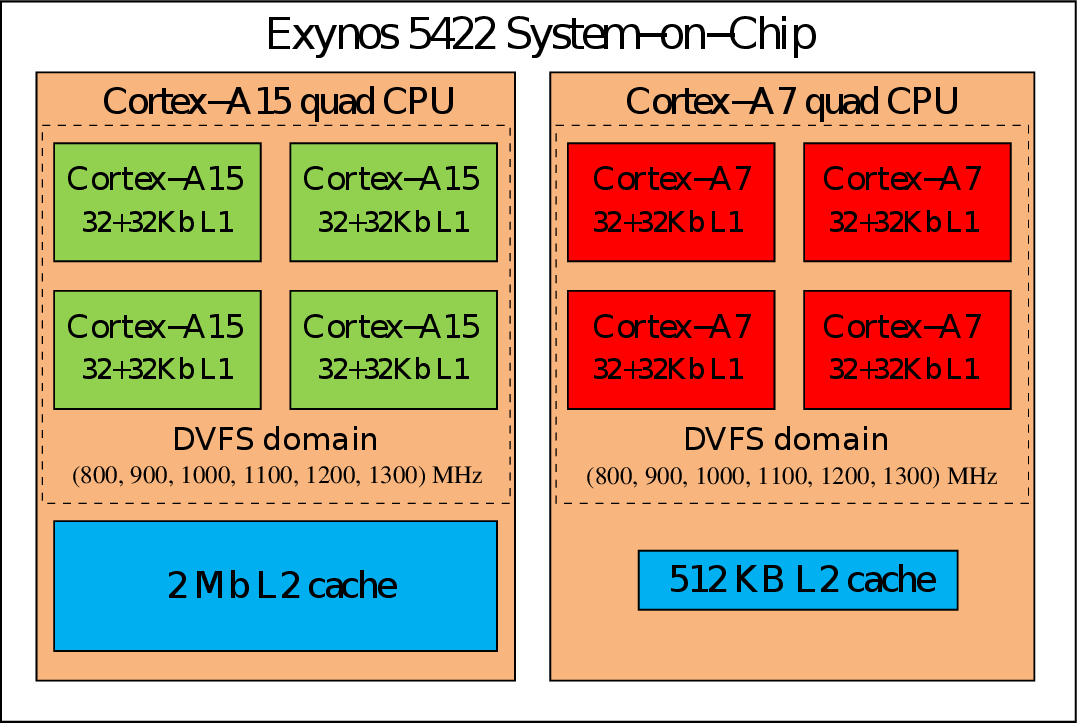}
  \caption{Samsung Exynos 5422 SoC employed in our experiments.}
  \label{s2:fig:exynos}
\end{figure}

The target architecture for our experiments is an
\mbox{ODROID-XU3} board comprising a Samsung Exynos 5422 SoC
with an 32-bit ARM processor and 2GB DDR3 RAM. The chip features an ARM
Cortex-A15 quad-core processing cluster and a Cortex-A7 quad-core
processing cluster. Each ARM core (either Cortex-A15 or Cortex-A7) has a
\mbox{32+32-KByte L1} (instruction+data) cache. The four A15 cores share a
\mbox{2-MByte L2} cache, while the four A7 cores share a smaller 512-KByte
L2 cache. All cores of the same cluster share the same frequency of operation, clocking
from 800MHz to 1300MHz in steps of 100MHz in both cases.  
The board exposes independent power measurements for each cluster.
Figure~\ref{s2:fig:exynos} shows a schematic view of the Exynos SoC.


\section{Proposed energy-aware policies}
\label{sec:policies}

We introduce two different general approaches that pursue an improvement on
the energy efficiency of task-parallel codes on asymmetric architectures.
The first group of policies (named as \Pfs) is based on the dynamic application of DVFS techniques
at runtime.
The goal is to integrate these techniques on an 
asymmetry-aware scheduler, and to reduce energy consumption
by modifying the frequency of one of the clusters based on the internal state
of the scheduler, without further modifications on the scheduling algorithm. 
Pursuing the same goal, the second
group of policies (named \Pts) implements different asymmetry-aware scheduling
algorithmic variations on existing task schedulers.



\subsection{Policies based on frequency scaling (\Pfs)}

Applying DVFS techniques to a task-parallel problem requires three main
runtime decisions to be made, namely: (a) which frequencies (among those available) 
to use; (b) at which moments of the parallel execution these changes need to be made;
and (c) which elements of the architecture (among those that support DVFS) are
affected by the voltage/frequency scaling.
The set of frequencies that a processor can run at is usually defined by the
architecture, so the first decision is reduced to choosing between using
all the available frequencies or just a subset of them. The remaining decisions
are directly related to the specific problem to tackle, and the knowledge
that the task scheduler has of it.


Figure~\ref{s3:fig:FS:queues} shows, for a Cholesky factorization of a
$1024\times 1024$ matrix divided in blocks of dimension $64\times 64$,
the evolution in time of the amount of critical and non-critical tasks ready for execution ($N_{crit}$
and $N_{non\_crit}$, respectively, being $N_{ready} = N_{crit} + N_{non\_crit}$), together with the ratio 
between them ($R_{c\_nc} = N_{crit} / N_{non\_crit}$). In the following,
we also consider $N_{max}^{nc}$ and $N_{max}$ as the maximum amount of
ready non-critical tasks and ready tasks (critical and non-critical) observed from the beginning of the execution at each moment. 
Both values, $N_{max}^{nc}$ and $N_{max}$, are constantly monitored and updated at runtime by the scheduler.
Finally, $R_{non\_crit} = N_{non\_crit} / N_{max}^{nc}$ denotes
the ratio of non-critical ready tasks compared with the maximum amount observed
for this value.

\subsubsection{Policy \Pone. Tasks limited by the critical path}

Runtime task schedulers annotate tasks while the DAG is built and, typically, no
further external information is used; thus, it is possible that multiple paths of the DAG
are detected as critical at the same time. On an asymmetry-aware
scheduler like \BOTLEV, this situation entails that most of the tasks will be
executed by \BIG cores (as they are annotated as critical), while
\LITTLE cores will be in idle state until new non-critical tasks are
detected.
%
%
Asymmetry-aware task schedulers alleviate these situations by 
allowing critical tasks to be executed by both types of cores
until new non-critical tasks are ready to run. However, using \LITTLE cores to
execute critical tasks can slow down the execution as, despite the
fact that tasks can start their execution earlier due to the greater number
of available cores, running a task on a slow core can increase its execution time
meaningfully.

Our approach to respond to this situation is different, as is our goal (reducing energy
consumption): 
the \Pone policy leverages these moments --where the number of ready critical
tasks is greater than the number of ready non-critical tasks-- to 
reduce power consumption by decreasing the frequency of the \LITTLE
cluster. The side effect is that the execution
time of non-critical tasks increases, but as the global execution time is
limited by the critical tasks executed on the \BIG cluster, delaying the
execution of non-critical tasks on these moments should not dramatically impact 
the global performance.

In \Pone, the decision on which frequency the \LITTLE cluster should run at is made
by the scheduler each time the number of ready tasks changes (i.e., when a task becomes
ready or a ready task is executed by an idle core), and it is based on the
{\em relation} between the sizes of both queues ($R_{c\_nc}$), that determines the specific
frequency step that will be applied to the \LITTLE cluster.
For example, if $R_{c\_nc} == 2$, 
the \LITTLE cluster will run at its second maximum available frequency (in this case, 1200MHz); if $R_{c\_nc} == 5$,
the cluster will run at its fifth maximum frequency available (in this case, 900MHz).

\begin{figure}[!t]
\subfloat[Evolution of the number of critical and non-critical ready tasks.]{\includegraphics[width=0.96\columnwidth]{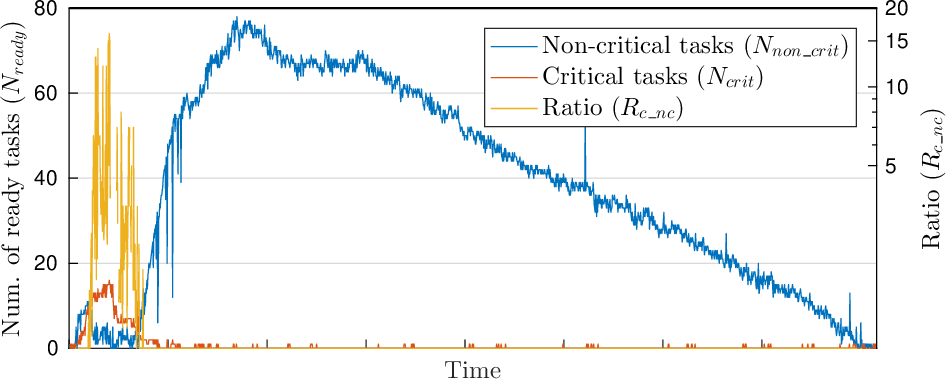}%
\label{s3:fig:FS:queues}}
\hfil
\subfloat[Policy \Pone.]{\includegraphics[width=0.97\columnwidth]{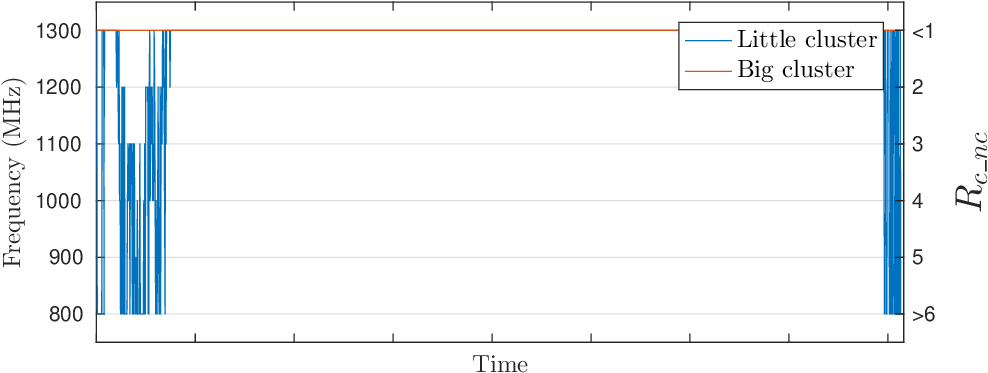}%
\label{s3:fig:FS:P1}}
\hfil
\subfloat[Policy \Ptwo. Notice that policy \Pthree will have the same behavior,
but applied to the other cluster.]{\includegraphics[width=1.0\columnwidth]{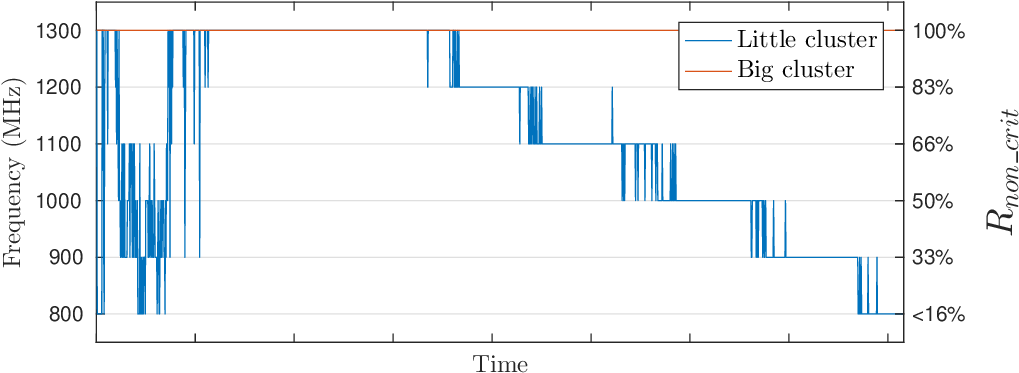}%
\label{s3:fig:FS:P2}}
\hfil
\subfloat[Policy \PtwoP.]{\includegraphics[width=1.0\columnwidth]{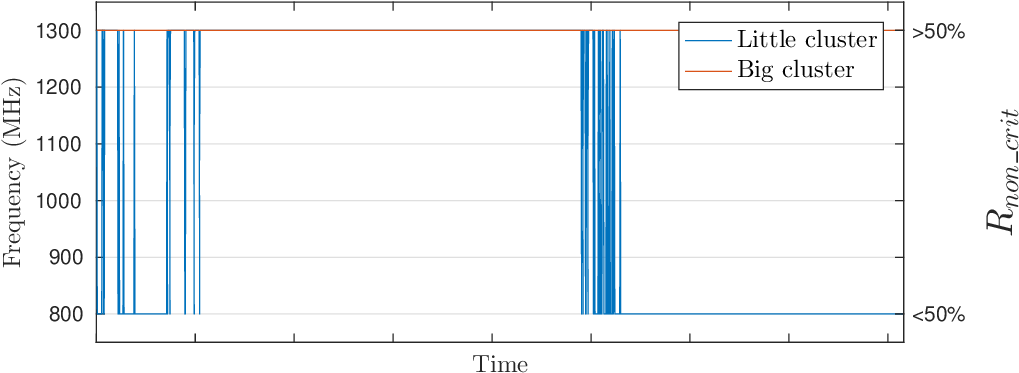}%
\label{s3:fig:FS:P22}}

\caption{Behavior of each \Pfs policy when is applied to a Cholesky
  factorization of a $1024\times 1024$ matrix divided in blocks of
  $64\times 64$ elements.}
\label{s3:fig:FS}
\end{figure}


  Figure~\ref{s3:fig:FS:P1} reports the instantaneous frequency applied by the
  task scheduler when applying \Pone on the same execution as that shown in Figure~\ref{s3:fig:FS:queues}.
  Observe how, when the number of ready critical tasks is higher
  than the number of ready non-critical tasks (e.g. at the beginning and end stages
  of the execution in this example),
  the frequency of the \LITTLE cluster is scaled down, and how the frequency
  chosen for the cluster is directly related with $R_{c\_nc}$. Also, note how, when 
  $N_{non\_crit}$ increases, the policy forces the \LITTLE cores to
  run at a higher (even at the maximum) frequency.


\subsubsection{Policies \Ptwo and \PtwoP. \LITTLE cluster frequency scaled based
  on the workload}

Instead of modifying the frequency based on the {\em ratio} between the number
of both types of ready tasks, policies \Ptwo and \PtwoP modify the
frequency based on the {\em absolute amount} of non-critical tasks at each moment, i.e.,
if there is a high number of non-critical tasks, the \LITTLE cluster will
run at a high frequency, and if the number is low, the frequency will be
lower. 

In order to determine when the number of non-critical tasks is considered high or low,
$N_{non\_crit}$ is compared with $N_{max}^{nc}$. If higher,
\Ptwo and \PtwoP will consider that the number of 
non-critical tasks is high, and the \LITTLE cluster will run at its maximum frequency; 
if not, frequency is scaled down depending on the value of $R_{non\_crit}$.

The difference between \Ptwo and \PtwoP is the set of frequencies 
to select: while \Ptwo chooses one between all the available frequency steps according to 
$R_{non\_crit}$ (see Figure~\ref{s3:fig:FS:P2}),  
\PtwoP only uses the highest and lowest available frequencies (see Figure~\ref{s3:fig:FS:P22}). 
In this case, if the current number of non-critical tasks is lower than the 50\% of the maximum amount
recorded (that is, if $R_{non\_crit} < 0.5$), the frequency will be the lowest available, in other case, it will be the
highest.

{

Observing the evolution of $N_{crit}$ and $N_{non\_crit}$ 
in Figure~\ref{s3:fig:FS:queues}, two different phases can be distinguished: a first
phase where the number of ready non-critical tasks increases, and a second
phase where it decreases. This behaviour matches with a Cholesky
factorization \DAG, which enlarges very fast at the beginning, and it
reduces slowly later.  While the first phase occurs, the maximum amount of
ready non-critical tasks is growing, so the frequency which \LITTLE cluster
is running at is its maximum frequency; during the second phase,
the scheduler scales down frequency based on the amount of non-critical tasks and
available frequencies.
}

\subsubsection{Policy \Pthree. \BIG cluster frequency scaled based on the workload}

The behavior of policy \Pthree is similar to that of \Ptwo, but,
instead of modifying the frequency of the \LITTLE cluster, \Pthree scales the
frequency of the \BIG cluster.



\subsection{Policies based on task scheduling (\Pts)}

The \Pts policies described next are based on the same ideas as \Pfs policies
but, instead of applying DVFS techniques, they decide at runtime the phases in
which both clusters are considered to execute tasks, or just one of them
is used as a scheduling target. 
On one hand, using only one of the clusters in specific moments means that
power consumption is likely to decrease, but on the other hand, performance 
will also be affected. Our goal is to find a trade-off between both parts, and
thus to improve energy efficiency.

\subsubsection{Policies \Pfour and \Pfive. Making cluster unusable depending on the
  workload}

Similar to policies \Ptwo and \Pthree, these policies track the value of $N_{ready}$
at each moment, and determine when the amount of tasks is increasing
or decreasing (comparing this value with $N_{max}$). 
If the number of ready tasks is low enough, the policy will not assign any new
task to one of the clusters, making it to be in an idle state from the scheduler's perspective, and saving
power consumption. If the number of tasks increases later, the cluster
becomes available again and it will execute new tasks as they become available. 
The amount of tasks (or threshold) that determines when to disable or enable the cluster 
(denoted as $N_{thres}$ in the following) is configurable and not defined by the policy; several 
experiments with different values for $N_{thres}$ can be found in the next section.

The difference between policies \Pfour and \Pfive is that, while \Pfour
acts on the \LITTLE cluster, \Pfive acts disabling and enabling the \BIG
cluster. As \Pfive disables the \BIG cluster in some moments of the execution,
critical tasks are executed on \LITTLE cores until the \BIG cluster is
enabled again.

\begin{figure}
  \centering
  \begin{framed}
    \subfloat[Execution trace per core.]{\includegraphics[width=1.0\columnwidth]{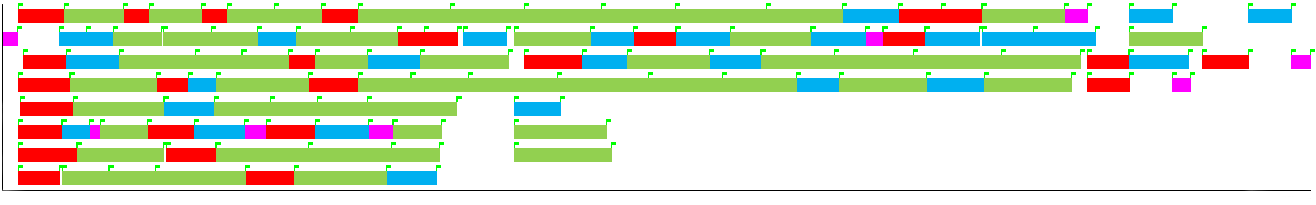}}
    \hfil
    \subfloat[Num. of ready tasks.]{\includegraphics[width=1.0\columnwidth]{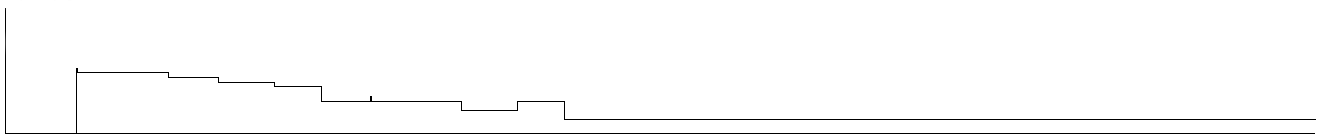}}

  \end{framed}

  \caption{Policy \Pfive: task scheduling based on the number of ready tasks,
    for a Cholesky factorization of a square $4096\times4096$ elements
    matrix, grouped in square blocks of $512\times512$ elements each
    executed on an \ODROID platform. 
    Color key: red=\TRSM, pink=\POTRF, blue=\SYRK, green=\GEMM, white=\IDLE.}
  \label{fig:s3:P5_evolution}
\end{figure}

Figure~\ref{fig:s3:P5_evolution} shows an execution of policy \Pfive
applied to a Cholesky factorization, where the cluster is disabled when the
current number of ready tasks is under the 30\% of $N_{max}$ (that is, $N_{thres} = 30\%$). Each
line in the trace corresponds to a specific core executing tasks (coloured areas)
or in idle state (white areas).
The trace has been obtained on an \ODROID platform, where cores (numbered
from the top to the bottom) 0-3 belong to the \LITTLE cluster, and cores 4-7 to the \BIG
cluster. The plot at the bottom shows the number of ready tasks at each
moment. 
Observe how, at the beginning, the task scheduler assigns tasks to
all the available cores, until the number of ready tasks is under 30\% of
maximum recorded; from that moment on, no tasks are assigned to \BIG cluster.
As there are less cores to execute ready tasks, in some
moments of the execution the number of ready tasks becomes greater than $N_{thres}$,
starting \BIG cores to execute ready tasks until number of ready tasks decreases
again and the cluster becomes unavailable for scheduling purposes. 

\subsubsection{Policy \Psix. Cluster disabled based on workload}

\begin{figure}
  \centering
  \includegraphics[width=1.0\columnwidth]{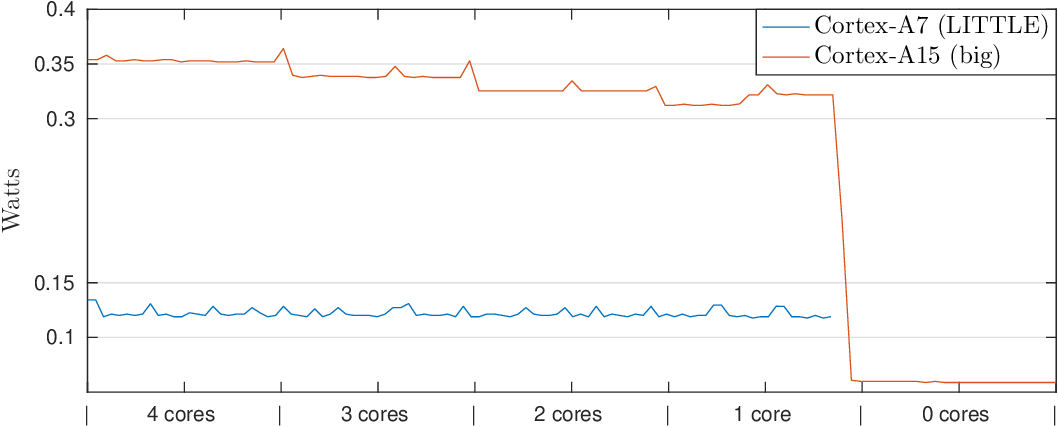}
  \caption{Power consumption of each cluster on idle state with different
    number of active cores. Linux kernel does not allow switching off the
    whole \LITTLE cluster, thus measures could not be made for this scenario.}
  \label{fig:s3:apagado_Odroid}
\end{figure}

Some platforms allow switching
off one of the clusters under demand via the OS, which entails an important 
decrease on power consumption, as shown in Figure~\ref{fig:s3:apagado_Odroid}. 
Policy \Psix is similar to
policy \Pfive, but in addition to deactivating the \BIG cluster to the task
scheduler, it switches it off completely\footnote{As the Linux Kernel does not allow powering off the
core number zero in our platform, experiments related with switching off the \LITTLE
cluster could not be performed.}.

\section{Experimental results}
\label{sec:results}

In the following, we report the experimental results obtained for the Cholesky
factorization on the \ODROID SoC applying the proposed policies. In all cases,
we show results for performance (in terms of GFLOPS), average power (in Watts)
and energy efficiency (in GFPLOPS/Watt). All experiments were carried out
using single precision and gathering power results from the internal meters in
the board. Each experiment was repeated ten times, showing the average
measurements in the following.

\subsection{Policies based on frequency scaling (\Pfs)}
\begin{figure}
  \centering
  \includegraphics[width=\columnwidth]{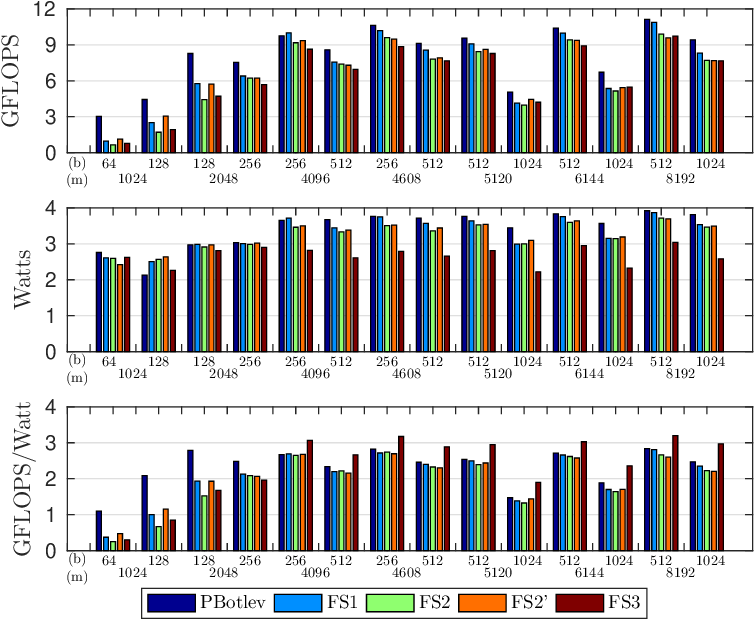}
  \caption{Experimental measures for policies from \Pone to \Pthree on an \ODROID
    platform. Policy \Pzero stands for a normal execution using the
    asymmetry-aware scheduler \BOTLEV without any policy. Tags in the
    horizontal axis represent the sizes of the matrix and blocks of each
    experiment.}
  \label{fig:s4:p1_p3-cholesky-odroid}
\end{figure}

Figure~\ref{fig:s4:p1_p3-cholesky-odroid} shows the results obtained when
applying policies from \Pone to \Pthree to different Cholesky factorizations on an
\ODROID platform. The experiments cover a range of different matrix sizes
and block dimensions. 
%
A number of general, preliminar remarks can be extracted from the results. Depending on the
matrix size, the conclusions differ, namely:

\begin{itemize}
\item For small matrices, ($m \le 2048$), there is a considerable difference between
  the performance obtained when the factorization is made without any
  policy (named \Pzero in the Figures) and when using any of our
  policies. This big difference in the performance has a huge impact on the
  energy efficiency.

\item For large matrices ($m \ge 4096$), applying
  our policies also implies a penalty in performance, as expected. However, energy
  efficiency measurements are very similar to \Pzero. In this case, \Pthree clearly outperforms
  \Pzero in terms of energy efficiency. In addition, as a positive side effect and for this range of problem sizes, 
  the application of any \Pfs policy clearly reduces the average power consumption (in Watts) of
  the execution.

\end{itemize}


\begin{table}
  \centering
  \caption{Improvement of average power consumption (in Watts) for policies
    from \Pone to \Pthree.}
  \label{tab:s4:mejora-potencia-P1_P3}
  {\scriptsize
    \setlength{\tabcolsep}{2.5pt}
    \begin{tabular}{lccccccccccccc}
      \hline\noalign{\smallskip}
     \phantom{a} & \multicolumn{13}{c}{Matrix size ($m\times m$) and block size ($b\times b$).} \\ \hline
      \multicolumn{1}{r}{\texttt{(m)}} & \multicolumn{2}{c}{1024} & \phantom{a} & \multicolumn{2}{c}{4096} & \multicolumn{2}{c}{4608} & \multicolumn{2}{c}{5120} & \multicolumn{2}{c}{6144} & \multicolumn{2}{c}{8192} \\
      \multicolumn{1}{r}{\texttt{(b)}} & 64 & 128 & \phantom{.} & 256 & 512 & 256 & 512 & 512 & 1024 & 512 & 1024 & 512 & 1024 \\ \hline\noalign{\smallskip}

{\sc \Pone} & \br{-0.93} & \br{-0.30} & \phantom{.} & \br{-0.04} & \fg{0.17} & \fg{0.00} & \fg{0.13} & \fg{0.11} & \fg{0.42} & \fg{0.07} & \fg{0.39} & \fg{0.05} & \fg{0.27} \\
{\sc \Ptwo}  & \br{-1.22} & \br{-0.46} & \phantom{.} & \fg{0.15} & \fg{0.27} & \fg{0.22} & \fg{0.30} & \fg{0.21} & \fg{0.41} & \fg{0.22} & \fg{0.40} & \fg{0.20} & \fg{0.34} \\
{\sc \PtwoP}  & \br{-0.78} & \br{-0.09} & \phantom{.} & \fg{0.13} & \fg{0.22} & \fg{0.19} & \fg{0.23} & \fg{0.20} & \fg{0.33} & \fg{0.18} & \fg{0.35} & \fg{0.22} & \fg{0.31} \\
{\sc \Pthree} & \br{-1.09} & \br{-0.21} & \phantom{.} & \fg{0.73} & \fg{0.93} & \fg{0.87} & \fg{0.96} & \fg{0.89} & \fg{1.18} & \fg{0.85} & \fg{1.21} & \fg{0.86} & \fg{1.21} \\
\noalign{\smallskip}\hline
    \end{tabular}
  }
\end{table}

Diving into details of average power and energy efficiency
results for each policy, a number of specific insights can be extracted.
First, the gap in performance, average power and energy efficiency between policies \Ptwo and
\PtwoP is not remarkable and, similar to policy \Pone, experimental results do not show any improvement 
in terms of energy efficiency when using these
policies for this application and platform. However, 
a decrease in the power consumption is observed, making these policies of great appeal
when targeting environments where the power consumption is limited by
design. Table~\ref{tab:s4:mejora-potencia-P1_P3} reports the decrease of
power consumption (in Watts) achieved for each policy. In the first set of
matrices (the ones with lowest size), the power consumption increases, but,
in the second group, the power consumption decreases in all matrix
configurations and for all the policies, achieving a decrease up to 0.41 Watts (12.85\%)
for policies \Ptwo and \PtwoP, and a decrease up to 1.21 Watts (34.88\%) for policy \Pthree.

Second, the penalty introduced by the application of \Pone, \Ptwo and \PtwoP 
in terms of performance does not make up for the improvements in average power 
introduced by the frequency scaling in those policies. Thus, for this problem, 
they actually increase the energy consumption of the solution.

Finally, from Figure~\ref{fig:s4:p1_p3-cholesky-odroid} we can observe that the
policy which obtains the best results is \Pthree, outperforming \BOTLEV
in terms of energy efficiency. Table~\ref{tab:s4:mejora-gflopsw-P1_P3} reports a detailed study of the energy efficiency
improvement (in \mbox{GFLOPS/Watt}) of each policy and matrix configuration
compared with a normal execution using \BOTLEV. For matrices larger than
2048 elements, \Pthree obtains a rise on energy efficiency, achieving
improvements from 11.7\% up to 29.3\%.



\begin{table}
  \centering
	\caption{ Improvement of energy efficiency (in GFLOPS/Watt) for
          policies \Pone-\Pthree of different configurations of a Cholesky
          factorization on an \ODROID platform.}
  \label{tab:s4:mejora-gflopsw-P1_P3}

  {\scriptsize
    \setlength{\tabcolsep}{1.5pt}
    \begin{tabular}{lccccccccccccc}
      \hline\noalign{\smallskip}
      \phantom{a} & \multicolumn{13}{c}{Matrix size ($m \times m)$ and block size ($b \times b$).} \\ \hline
      \multicolumn{1}{r}{\texttt{(m)}} & \multicolumn{2}{c}{1024} & \phantom{a} & \multicolumn{2}{c}{4096}& \multicolumn{2}{c}{4608} & \multicolumn{2}{c}{5120} & \multicolumn{2}{c}{6144} & \multicolumn{2}{c}{8192} \\
      \multicolumn{1}{r}{\texttt{(b)}} & 64 & 128 & \phantom{.} & 256 & 512 & 256 & 512 & 512 & 1024 & 512 & 1024 & 512 & 1024 \\ \hline\noalign{\smallskip}

      {\sc \Pone} & \br{-4.83} & \br{-3.84} & \phantom{a} & \fg{0.04} & \br{-0.19} & \br{-0.12} & \br{-0.07} & \br{-0.05} & \br{-0.10} & \br{-0.06} & \br{-0.20} & \br{-0.03} & \br{-0.12} \\
      {\sc \Ptwo} & \br{-5.12} & \br{-4.87} & \phantom{a} & \br{-0.04} & \br{-0.16} & \br{-0.11} & \br{-0.16} & \br{-0.17} & \br{-0.16} & \br{-0.11} & \br{-0.26} & \br{-0.18} & \br{-0.25} \\
      {\sc \PtwoP} & \br{-4.64} & \br{-2.49} & \phantom{a} & \br{-0.01} & \br{-0.23} & \br{-0.16} & \br{-0.19} & \br{-0.12} & \br{-0.04} & \br{-0.15} & \br{-0.19} & \br{-0.25} & \br{-0.27} \\
      {\sc \Pthree} & \br{-5.01} & \br{-4.22} & \phantom{a} & \fg{0.41} & \fg{0.31} & \fg{0.34} & \fg{0.43} & \fg{0.41} & \fg{0.43} & \fg{0.31} & \fg{0.48} & \fg{0.36} & \fg{0.50} \\
      \noalign{\smallskip}\hline
    \end{tabular}
  }
\end{table}

\subsection{Policies based on task scheduling (\Pts)}

\begin{table}
  \centering
  \caption{Amount of time when the \LITTLE cluster is unusable for
    different configurations of policy \Pfour (rows) and Cholesky factorization
    (columns) in a \JUNO platform.}
  \label{tab:s4:tam-colas-tiempo}
  {\scriptsize
    \setlength{\tabcolsep}{1.5pt}
    \begin{tabular}{lccccccccccccc}
      \hline\noalign{\smallskip}
      \phantom{a} & \multicolumn{13}{c}{Matrix size \texttt{($m \times m$)} and block size ($b \times b$).} \\ \hline
      \multicolumn{1}{r}{\texttt{(m)}} & \multicolumn{2}{c}{1024} & \phantom{a} & \multicolumn{2}{c}{4096} & \multicolumn{2}{c}{4608} & \multicolumn{2}{c}{5120} & \multicolumn{2}{c}{6144} & \multicolumn{2}{c}{8192} \\
      \multicolumn{1}{r}{\texttt{(b)}} & 64 & 128 & \phantom{.} & 256 & 512 & 256 & 512 & 512 & 1024 & 512 & 1024 & 512 & 1024 \\ \hline\noalign{\smallskip}

      {\sc 50\%} & 69.36 & 45.83 & \phantom{.} & 43.44 & 50.83 & 39.39 & 38.48 & 40.91 & 42.07 & 39.15 & 35.24 & 40.32 & 48.33 \\
      {\sc 40\%} & 68.20 & 31.25 & \phantom{.} & 29.41 & 33.33 & 30.61 & 32.73 & 32.73 & 30.54 & 32.83 & 30.42& 30.15 & 37.92 \\
      {\sc 30\%} & 63.42 & 34.58 & \phantom{.} & 21.14 & 32.50 & 20.88 & 24.85 & 25.00 & 28.06 & 23.08 & 25.43& 21.81 & 33.75 \\
      {\sc 20\%} & 20.40 & 17.92 & \phantom{.} & 11.40 & 31.67 & 11.97 & 17.27 & 18.41 & 21.04 & 14.56 & 19.12& 13.11 & 26.67 \\
      {\sc 10\%} & 23.10 & 15.00 & \phantom{.} & 5.27  & 20.00 & 4.87 & 11.52  & 9.55 & 15.13 &7.69 & 10.3 & 5.64 & 12.92 \\
      \noalign{\smallskip}\hline
    \end{tabular}
  }
\end{table}

Opposite to \Pfs, \Pts policies do not pre-define a specific
moment of the execution in which a cluster is disabled. 
The experiments described below take into account different configurations 
of the policies, from
disabling the cluster when the amount of ready tasks is 50\% of the maximum
amount recorded (that is, $N_{thres} = 50\%$), 
to disabling it when the amount is only at 10\%. 
Note that disabling the cluster when the current number of ready tasks is, for
example, half of the maximum amount recorded does not imply that the cluster will be
unusable 50\% of the execution time. 
Table~\ref{tab:s4:tam-colas-tiempo} shows the percentage of time in which
the \LITTLE cluster is unusable for policy \Pfour, depending on the configuration of
the policy and problem dimensions.

\begin{figure}
  \centering
  \includegraphics[width=\columnwidth]{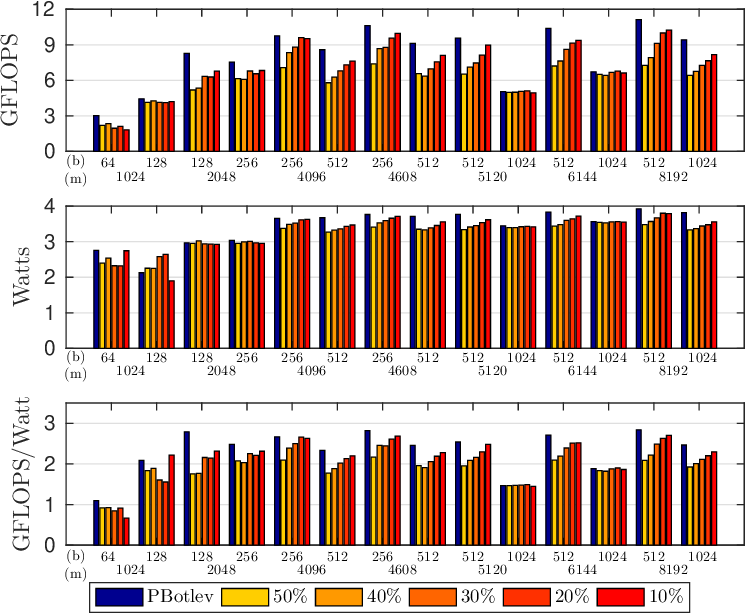}
  \caption{Experimental results for different \Pfour configurations applied to
    multiple matrix sizes.}
  \label{fig:s4:p4-cholesky-odroid}
\end{figure}

\begin{figure}
  \centering
  \includegraphics[width=\columnwidth]{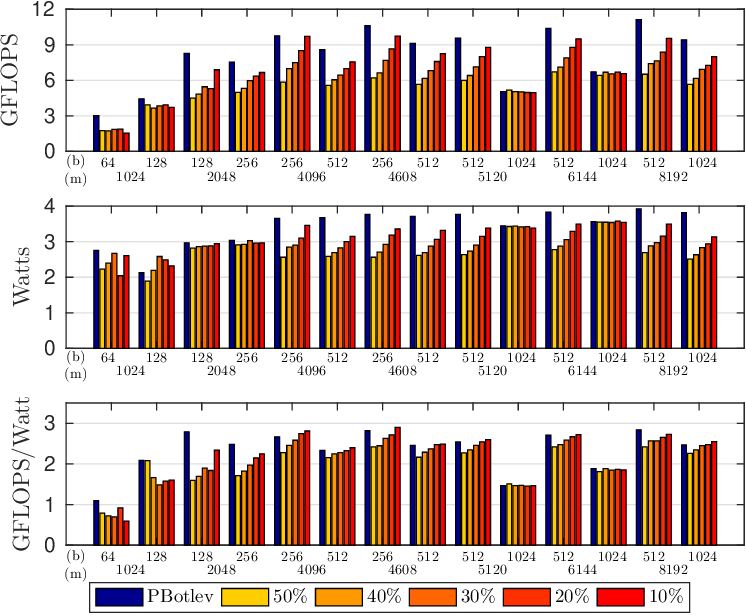}
  \caption{Experimental results for different \Pfive configurations applied to
    multiple matrix sizes.}
  \label{fig:s4:p5-cholesky-odroid}
\end{figure}

Figures~\ref{fig:s4:p4-cholesky-odroid} and~\ref{fig:s4:p5-cholesky-odroid}
report the behavior of policies \Pfour and \Pfive, respectively, for different matrix
sizes and policy configurations, in terms of performance, average power and energy
efficiency. 
%
%
The experiments reveal that $N_{thres}$ has a high impact on the final
performance, independently of the cluster which is affected by the policy.
%
In general, both policies exhibit worse energetic results than not using any
policy. Whereas policy \Pfive has similar energy efficiency results than \Pzero, the results
obtained when \Pfour is used are worse than when not using it.



Table~\ref{tab:mejora-gflopsw-p5} shows the improvement of GFLOPS/Watt
obtained when policy \Pfive is compared with a normal execution (policy
\Pzero). Although this policy does not achieve an improvement in energy
efficiency, it obtains similar energy-efficiency measurements with lower
overall power consumption, making this policy, together with policies \Ptwo
and \PtwoP, good candidates for scenarios where the power consumption is
limited.

\begin{table}
  \centering
  \caption{Improvement of energy efficiency for policy \Pfive.}
  \label{tab:mejora-gflopsw-p5}
  {\scriptsize
    \setlength{\tabcolsep}{2.0pt}
    \begin{tabular}{lccccccccccccc}
      \hline\noalign{\smallskip}
      \phantom{a} & \multicolumn{13}{c}{Matrix size ($m \times m$) and block size ($b \times b$).} \\ \hline
      \multicolumn{1}{r}{\texttt{(m)}} & \multicolumn{2}{c}{1024} & \phantom{.} & \multicolumn{2}{c}{4096} & \multicolumn{2}{c}{4608} & \multicolumn{2}{c}{5120} & \multicolumn{2}{c}{6144} & \multicolumn{2}{c}{8192} \\
      \multicolumn{1}{r}{\texttt{(b)}} & 64 & 128 & \phantom{.} & 256 & 512 & 256 & 512 & 512 & 1024 & 512 & 1024 & 512 & 1024 \\ \hline\noalign{\smallskip}

{\sc 10\%} & \br{-3.67} & \fg{4.07} & \phantom{.} & \br{-0.52} & \br{-0.27} & \br{-0.52} & \br{-0.37} & \br{-0.32} & \fg{0.05} & \br{-0.33} & \br{-0.08} & \br{-0.44} & \br{-0.22} \\               
{\sc 20\%} & \br{-3.98} & \fg{1.98} & \phantom{.} & \br{-0.30} & \br{-0.16} & \br{-0.48} & \br{-0.23} & \br{-0.25} & \br{-0.01} & \br{-0.27} & \fg{0.00} & \br{-0.28} & \br{-0.13} \\
{\sc 30\%} & \br{-4.04} & \br{-0.99} & \phantom{.} & \br{-0.15} & \br{-0.12} & \br{-0.27} & \br{-0.13} & \br{-0.12} & \fg{0.00} & \br{-0.15} & \br{-0.03} & \br{-0.29} & \br{-0.03} \\ 
{\sc 40\%} & \br{-3.69} & \fg{1.56} & \phantom{.} & \fg{0.06} & \br{-0.06} & \br{-0.15} & \br{-0.01} & \br{-0.02} & \br{-0.02} & \br{-0.06} & \br{-0.01} & \br{-0.19} & \fg{0.01} \\
{\sc 50\%} & \br{-4.04} & \br{-1.47} & \phantom{.} & \fg{0.16} & \fg{0.04} & \fg{0.06} & \fg{0.01} & \fg{0.05} & \br{-0.01} & \fg{0.01} & \br{-0.03} & \br{-0.11} & \fg{0.08}\\
\noalign{\smallskip}\hline
    \end{tabular}
  }
\end{table}

\begin{figure}
  \centering
  \includegraphics[width=\columnwidth]{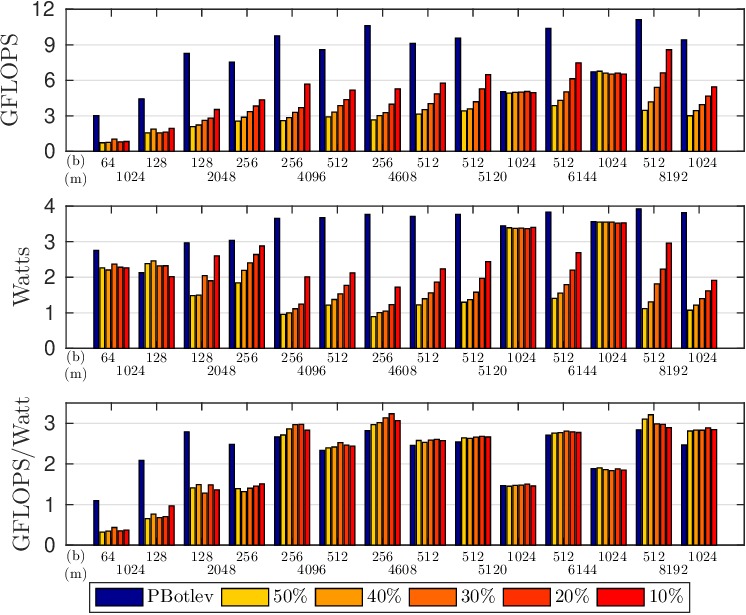}
  \caption{Experimental results for different \Psix configurations applied to
    multiple matrix sizes.}
  \label{fig:s4:p6-cholesky-odroid}
\end{figure}

Policy \Psix does achieve an improvement in terms of energy efficiency on
most of the tested configurations. Figure~\ref{fig:s4:p6-cholesky-odroid}
shows the results obtained when this policy was applied for different
problem dimensions. The application of the policy attains an improvement
of up to 17.1\%. Table~\ref{tab:mejora-gflopsw-p6} shows the improvements
for each configuration in terms of GFLOPS/Watt.

Although policies \Pfive and \Psix exhibit similar behavior (policy \Pfive does not
use \BIG cores meanwhile policy \Psix switches them off), the performance
obtained is lower for policy \Psix. This overhead is probably caused by the
OS when it migrates the processes running on a \BIG core to a
\LITTLE one when a complete cluster is switched off (and similarly when it
is switched on again). However, due to the considerable decrease in power
consumption when the cluster is off (as shown in
Figure~\ref{fig:s3:apagado_Odroid}), the decrease in performance does not
entail a big impact on the overall energy efficiency.

\begin{table}
  \centering
  \caption{Energy performance improvement (in GFLOPS/Watt) for different \Psix
    policy configurations compared with a normal execution using \BOTLEV
    (policy \Pzero).}
  \label{tab:mejora-gflopsw-p6}
  {\scriptsize
    \setlength{\tabcolsep}{2.0pt}
    \begin{tabular}{lccccccccccccc}
      \hline\noalign{\smallskip}
      \phantom{a} & \multicolumn{13}{c}{Matrix size ($m \times m$) and block size ($b \times b$).} \\ \hline
      \multicolumn{1}{r}{\texttt{(m)}} & \multicolumn{2}{c}{1024} & \phantom{.} & \multicolumn{2}{c}{4096} & \multicolumn{2}{c}{4608} & \multicolumn{2}{c}{5120} & \multicolumn{2}{c}{6144} & \multicolumn{2}{c}{8192} \\
      \multicolumn{1}{r}{\texttt{(b)}} & 64 & 128 & \phantom{.} & 256 & 512 & 256 & 512 & 512 & 1024 & 512 & 1024 & 512 & 1024 \\ \hline\noalign{\smallskip}

{\sc 10\%} & \br{-4.98} & \br{-5.02} & \phantom{.} & \br{-0.16} & \br{-0.08} & \br{-0.02} & \fg{0.01} & \fg{0.02} & \br{-0.02} & \br{-0.01} & \fg{0.02} & \fg{0.24} & \fg{0.33}\\
{\sc 20\%} & \br{-4.95} & \br{-4.70} & \phantom{.} & \fg{0.00} & \br{-0.05} & \fg{0.03} & \br{-0.03} & \fg{0.01} & \fg{0.01} & \fg{0.01} & \br{-0.02} & \fg{0.35} & \fg{0.35} \\
{\sc 30\%} & \br{-4.71} & \br{-4.83} & \phantom{.} & \fg{0.12} & \fg{0.07} & \fg{0.16} & \fg{0.04} & \fg{0.05} & \fg{0.01} & \fg{0.05} & \br{-0.05} & \fg{0.13} & \fg{0.33} \\
{\sc 40\%} & \br{-4.92} & \br{-4.44} & \phantom{.} & \fg{0.15} & \fg{0.02} & \fg{0.29} & \fg{0.07} & \fg{0.09} & \fg{0.03} & \fg{0.04} & \br{-0.01} & \fg{0.13} & \fg{0.41} \\
{\sc 50\%} & \br{-4.89} & \br{-3.92} & \phantom{.} & \fg{0.06} & \fg{0.02} & \fg{0.14} & \fg{0.06} & \fg{0.09} & \br{-0.01} & \fg{0.04} & \br{-0.03} & \fg{0.05} & \fg{0.37} \\
\noalign{\smallskip}\hline
    \end{tabular}
  }
\end{table}



\section{Conclusions}
\label{sec:conclusions}

In this paper we have explored a number of ways to extend an
asymmetry-aware scheduler to optimize the energy efficiency of
task-parallel applications, focusing on ARM \mbox{big.LITTLE}
systems-on-chip. 
From the observations made for an illustrative dense linear algebra application
with complex data dependencies among tasks (the Cholesky factorization), a number
of insights have been extracted, namely: (1) scaling the frequency of the \LITTLE cluster does
not have a positive effect on the energy efficiency, but a reduction in
average power consumption is constantly achieved; (2) scaling the frequency of the
\BIG cluster does achieve considerable improvements on energy efficiency, increasing it
up to 29.3\%; (3) we have demonstrated that disabling the use of one of the clusters in some
moments of the execution also achieves a decrease on power consumption, but
not in energy efficiency, unless the switching off of the whole cluster is
supported by the hardware and OS, with improvements on energy efficiency of up to 17.1\%.


While the Cholesky factorization is representative of common operations in the dense linear
algebra field, future work will extend the experimental study to applications with different
features, and also to include extended levels of heterogeneity (e.g. including low-power GPUs
in the SoC). Automatically predicting optimal policies for a given application/architecture
is also in our roadmap.

\section*{Acknowledgments}
This work has been  supported by the EU (FEDER) and the Spanish MINECO, 
under grants TIN 2015-65277-R, TIN 2012-32180 and FPU15/02050.




\bibliographystyle{IEEEtran}

\end{document}